\documentclass[prl,preprint,
  showpacs,showkeys,lengthcheck,
  nofootinbib,
  tightenlines,twocolumn,notitlepage,
  preprintnumbers,superscriptaddress]{revtex4-1}

\usepackage[utf8]{inputenc}
\usepackage{newtxtext}
\usepackage[mathcal]{euscript}

\usepackage{graphicx}
\usepackage[usenames,dvipsnames]{xcolor}
\graphicspath{{figures/}}

\usepackage{hyperref}
\hypersetup{colorlinks=true,citecolor=blue,linkcolor=blue,urlcolor=blue}

\usepackage{diagbox}
\usepackage{booktabs}

\usepackage{amsmath,amssymb,amsfonts}
\usepackage{slashed}
\usepackage{bm}

\begin{document}

\title{Genuine empirical pressure within the proton}

\author{Adam Freese}
\email{afreese@uw.edu}
\affiliation{Department of Physics, University of Washington, Seattle, WA 98195, USA}

\author{Gerald A. Miller}
\email{miller@uw.edu}
\affiliation{Department of Physics, University of Washington, Seattle, WA 98195, USA}

\begin{abstract}
  A phenomenological extraction of pressure within the proton has recently
  been performed using JLab CLAS data
  (arXiv:2104.02031~[nucl-ex]~\cite{Burkert:2021ith}).
  The extraction used a 3-dimensional Breit frame description
  in which the initial and final proton states have different momenta.
  Instead, we obtain the two-dimensional transverse light front
  pressure densities that incorporate relativistic effects
  arising from the boosts that cause the initial and final states to differ.
  The mechanical radius is then determined to be
  $0.518~ \pm 0.062_{\mathrm{fit}} \pm 0.126_{\mathrm{sys}}$~fm,
  which is smaller than the electric charge radius and larger than
  the light front momentum radius.
  The forces within the proton are shown to be predominantly repulsive
  at distances less than $0.43~\pm 0.12~\mathrm{fm}$ from the center,
  and predominantly attractive further out.
\end{abstract}

\preprint{NT@UW-21-03}

\maketitle


\section{Introduction}

The goal of determining the magnitude and spatial distribution of forces within hadrons
has garnered great recent interest~\cite{Polyakov:2018zvc,Shanahan:2018nnv,Burkert:2018bqq,Kumericki:2019ddg,Freese:2019bhb,Anikin:2019ufr,Neubelt:2019sou,Varma:2020crx}. 
Information about internal forces within hadrons is encoded in the
energy-momentum tensor (EMT)~\cite{Polyakov:2002yz,Polyakov:2018zvc,Freese:2021czn},
which additionally contains information about the decomposition and distribution
of energy via a form factor, $A(t)$~\cite{Ji:1994av,Ji:1995sv,Lorce:2017xzd,Hatta:2018sqd,Rodini:2020pis,Metz:2020vxd}
and angular momentum via a form factor, $J(t)$~\cite{Ashman:1987hv,Ji:1996ek,Leader:2013jra}.
The variable $t$ represents the square of the momentum transfer
between initial and final proton states.
The focus here is the third form factor,
$D(t)$~\cite{Polyakov:2018zvc},
which encodes information about internal forces.
The three form factors represent separately conserved contributions to the EMT.

Recently, data from Jefferson Lab have been used to infer $D(t)$ and
the pressure distribution within the proton~\cite{Burkert:2018bqq,Burkert:2021ith}.
The obtained three-dimensional pressure distribution does not incorporate
relativistic effects caused boosts
that must be incorporated when $t R^2\sim 1$,
where $R$ is a measure of the size of the system.
Obtaining spatial distributions requires an integral over all values of $t$,
so determining the proton's internal structure requires a fully relativistic approach.

The relativistic effects due to boosts can be incorporated into
spatial densities by using light front coordinates and defining
the density at fixed light front
time~\cite{Burkardt:2002hr,Miller:2007uy,Miller:2009sg,Miller:2018ybm,Freese:2021czn}.
This can be done because the Poincar\'e group has a Galilean subgroup
that commutes with the light front
Hamiltonian~\cite{Dirac:1949cp,Susskind:1967rg,Brodsky:1997de}.
The densities obtained in this way
involve integrating out a spatial coordinate
in the light front direction, giving a two-dimensional density
on the transverse plane.
The formalism for using light front coordinates to obtain a relativistically
correct pressure density was explicated in Ref.~\cite{Freese:2021czn}.
Thus  we use the light front formalism to obtain a
relativistically correct pressure density from the Jefferson Lab $D(t)$ extraction.


\section{Light front formalism}
\label{sec:formalism}

In the light front formalism, spacetime is parametrized in terms of coordinates
$(x^+,x^-,\mathbf{x}_\perp)$,
where ${x^\pm = \frac{1}{\sqrt{2}}\big(x^0\pm x^3\big)}$.
$x^+$ is considered the ``time'' variable.
For transverse densities in particular,
all dependence on $x^-$ is integrated out,
giving a ${(2+1)}$-dimensional picture
in terms of the transverse spatial coordinates $\mathbf{x}_\perp$.
Within this ${(2+1)}$-dimensional picture,
the EMT---when sandwiched within physical state kets---can be written:
\begin{align}
  \langle \Psi |
  T^{\mu\nu}_{\mathrm{LF}}(x)
  | \Psi \rangle
  =
  u^\mu(x)
  u^\nu(x)
  \varepsilon(x)
  +
  S^{\mu\nu}(x)
  \,.
\end{align}
Here, $x=(x^+,\mathbf{x}_\perp)$,
and $\mu$ and $\nu$ range only over $+,1,2$. 
The wave-packet state $|\Psi\rangle$ is a superposition of momentum eigenstates
such that the transverse position is well-defined.
The variable
$\varepsilon(x)$ is the $P^+$ (light front momentum) density,
and $u^\mu(x)$ encodes the flow of the hadron---which includes
not just motion of the quarks and gluons within it,
but also movement of the wave packet due to dispersion.
The tensor $S^{\mu\nu}(x)$ is the ``pure stress tensor,''
and corresponds to the spatial components of the EMT as
measured by a locally comoving observer.
(i.e., an observer who sees $u^\mu(x)=0$ at their current location).
It is the pure stress tensor that
encodes the distribution of pressure and shear forces in the hadron.

For a transversely localized state with definite light front helicity
(i.e., polarized in the $\hat{z}$ direction),
the pure stress tensor is related to $D(t)$ by a two-dimensional
Fourier transform:
\begin{align}
  \label{eqn:Sij}
  S^{ij}(\mathbf{x}_\perp)
  =
  \frac{1}{4P^+}
  \int \frac{\mathrm{d}^2\boldsymbol{\Delta}_\perp}{(2\pi)^2}
  \Big(
  \boldsymbol{\Delta}_\perp^i \boldsymbol{\Delta}_\perp^j
  -
  \boldsymbol{\Delta}_\perp^2 \delta^{ij}
  \Big)
  \notag \\
  D(-\boldsymbol{\Delta}_\perp^2)
  \,
  e^{-i\boldsymbol{\Delta}_\perp\cdot\mathbf{x}_\perp}
  \,.
\end{align}
It can be decomposed into an
isotropic pressure function $p(\mathbf{x}_\perp)$
and a shear stress function $s(\mathbf{x}_\perp)$:
\begin{align}
  \label{eqn:Sij:ps}
  S^{ij}(\mathbf{x}_\perp)
  =
  \delta^{ij} p(\mathbf{x}_\perp)
  +
  \left( \frac{x_\perp^ix_\perp^j}{x_\perp^2} - \frac{1}{2}\delta^{ij} \right)
  s(\mathbf{x}_\perp)
  \,.
\end{align}
Since these are two-dimensional transverse quantities,
the pressure has units of force/length instead of force/area.
The pure stress tensor also gives the net force density acting at any point
$\mathbf{x}_\perp$ within the hadron via:
\begin{align}
  \label{eqn:force}
  \mathbf{F}_\perp^j(\mathbf{x}_\perp)
  =
  - \nabla_i S^{ij}(\mathbf{x}_\perp)
  \,.
\end{align}
For an equilibrium system such as an isolated hadron:
\begin{align}
  \label{eqn:zero}
  \mathbf{F}_\perp(\mathbf{x}_\perp) = 0
\end{align}
identically.
This force-balance condition can be seen to follow from Eq.~(\ref{eqn:Sij}).


\subsection{Radial and tangential pressures}

Although the net force everywhere in the hadron is zero,
there is nonetheless a static anisotropic pressure that would be felt by
a hypothetical pressure gauge immersed within the hadron.
$S^{ij}(\mathbf{x}_\perp)$ in particular encodes such pressures as measured
by a gauge that is comoving along with the hadron flow encoded in
$u^\mu(\mathbf{x}_\perp)$.
The force that would be measured by such a gauge is given by:
\begin{align}
  \mathbf{F}_{\mathrm{gauge}}^j
  =
  \int_L \mathrm{d}l \,
  \hat{u}_i \,
  T^{ij}(\mathbf{x}_\perp)
  \,,
\end{align}
where $L$ is the one-dimensional surface of the gauge and $\hat{u}_i$
is an inward-facing unit normal vector to that surface.

By appropriately considering gauges in different orientations,
one can obtain expressions for the radial and tangential pressure
within a hadron:
\begin{align}
  p_r(\mathbf{x}_\perp)
  &=
  \hat{r}_i \hat{r}_j T^{ij}(\mathbf{x}_\perp)
  =
  p(\mathbf{x}_\perp) + \frac{1}{2} s(\mathbf{x}_\perp)
  \\
  p_t(\mathbf{x}_\perp)
  &=
  \hat{\phi}_i \hat{\phi}_j T^{ij}(\mathbf{x}_\perp)
  =
  p(\mathbf{x}_\perp) - \frac{1}{2} s(\mathbf{x}_\perp)
  \,.
\end{align}
We follow Ref.~\cite{Lorce:2018egm} in calling these ``pressures.''
Refs.~\cite{Polyakov:2018zvc,Freese:2021czn} refer to $p_r(\mathbf{x}_\perp)$
(or its Breit frame analogue) as a ``normal force,''
but we avoid such nomenclature here in order to maintain clarity that
the net force everywhere in the hadron is zero.
Refs.~\cite{Polyakov:2018zvc,Lorce:2018egm,Freese:2021czn}
postulate $p_r(x_\perp) \geq 0$ as a stability condition,
but there are no sign constraints on $p_t(x_\perp)$.

Since $p_r(\mathbf{x}_\perp)$ is strictly non-negative,
it can be used to define a mechanical radius~\cite{Polyakov:2018zvc,Freese:2021czn},
which comes out as:
\begin{align}
  \label{eqn:radius:mechanical}
  \langle x_\perp^2(\mathrm{mech}) \rangle
  =
  \frac{
    \int \mathrm{d}^2 \mathbf{x}_\perp \,
    \mathbf{x}_\perp^2 p_r(\mathbf{x}_\perp)
  }{
    \int \mathrm{d}^2 \mathbf{x}_\perp \, p_r(\mathbf{x}_\perp)
  }
  =
  \frac{
    4D(0)
  }{
    \int_{-\infty}^0 \mathrm{d}t \, D(t)
  }
  \,.
\end{align}
This means that determining the mechanical radius requires
knowing $D(t)$ for both small and large values of $-t$. 


\section{Empirical transverse pressures}

The form factor $D(t)$ can in principle be extracted directly from the
Compton form factor $\mathcal{H}(\xi,t)$ using
dispersion relations~\cite{Diehl:2007jb,Anikin:2007tx,Pasquini:2014vua,Burkert:2018bqq}.
This was done in Ref.~\cite{Burkert:2018bqq} through a dispersive analysis of
deeply virtual Compton scattering data from
CLAS at Jefferson Lab~\cite{Girod:2007aa,Jo:2015ema,Burkert:2018bqq}.

A major caveat attached to the extraction
is that it includes only quarks,
since gluons do not contribute to DVCS at leading order.
Thus any pressure densities presented in this work are just
the quark contributions to these pressures.
Moreover, since gluons are not being included,
there is in principle
an additional form factor $\bar{c}_q(t)$ that can contribute
to the isotropic pressure density~\cite{Polyakov:2018zvc}.
However, phenomenological estimates~\cite{Lorce:2018egm,Hatta:2018sqd}
have $|\bar{c}_q(0)| \ll |D_q(0)|$,
so we neglect the contributions of this largely unknown form factor.

\begin{table}[t]
  \caption{
    Parameters for Eq.~(\ref{eqn:Dt}) from Ref.~\cite{Burkert:2021ith},
    including fit and systematic errors.
  }
  \label{tab:params}
  \setlength{\tabcolsep}{0.33em}
  \renewcommand{\arraystretch}{1.3}
  \begin{tabular}{@{} ccc @{}} 
    \toprule
    $ D(0) $ &
    $ \Lambda^2 $ (GeV$^2$) &
    $ \alpha $ \\
    \hline
    $          -1.47 \pm 0.06 \pm 0.14$ &
    $\phantom{-}1.02 \pm 0.13 \pm 0.21$ &
    $\phantom{-}2.76 \pm 0.23 \pm 0.48$ \\
    \bottomrule
  \end{tabular}
\end{table}

With these caveats in mind, the authors of Ref.~\cite{Burkert:2021ith}
have fit $D(t)$ to the following functional form:
\begin{align}
  \label{eqn:Dt}
  D(t)
  =
  \frac{D(0)}{(1-t/\Lambda^2)^\alpha}
  \,,
\end{align}
where $D(0)$, $\Lambda^2$, and $\alpha$ are the fit parameters.
The values obtained by the authors of Ref.~\cite{Burkert:2021ith}
are given in Table~\ref{tab:params}.

\begin{figure}
  \includegraphics[width=\columnwidth]{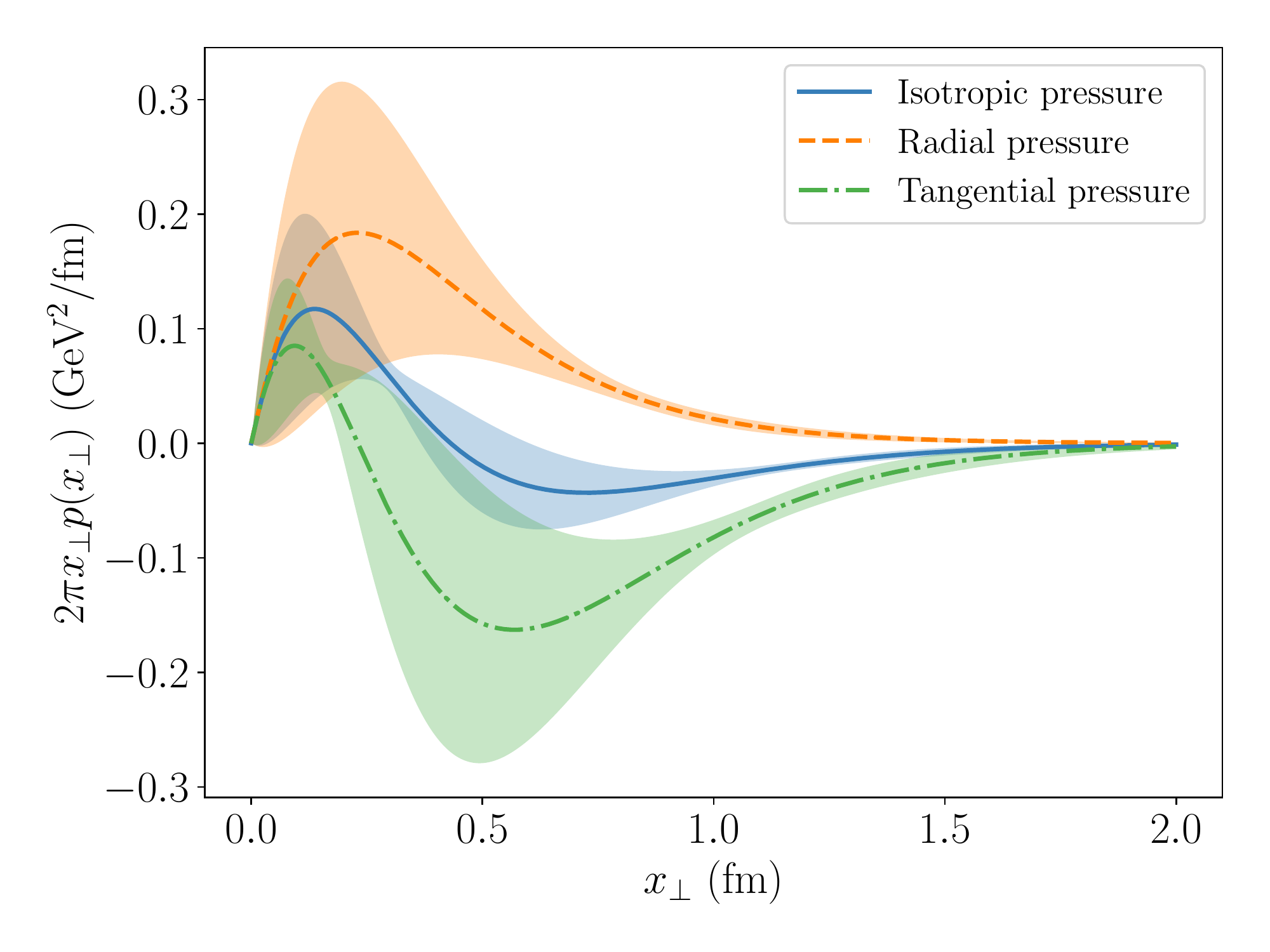}
  \caption{
    The isotropic, radial, and tangential pressures
    within the proton on the light front,
    as suggested by the parameters in Table~\ref{tab:params}.
    A state- and frame-dependent factor $1/P^+$ has been removed from the
    plotted quantities.
    The error band includes only fit uncertainty.
  }
  \label{fig:pressure}
\end{figure}

Using the empirical fit parameters for $D(t)$,
as well as the formalism explicated above,
we obtain empirical estimates for the isotropic, radial, and tangential pressure
of the proton in a definite light front helicity state.
For such a state, these pressures are function
of only the magnitude $x_\perp$ of $\mathbf{x}_\perp$.
These quantities, weighted by $2\pi x_\perp$,
are given in Fig.~\ref{fig:pressure}.
A factor of $1/P^+$ was removed from the plotted quantities,
making the plotted quantities state- and frame-independent.

The quantity $P^+ p(x_\perp)$ can also be cast into units of Pascals.
However, the numbers obtained should not be interpreted as literal forces
per unit area, since---by integrating out $x^-$---the light front formalism
is inherently $(2+1)$-dimensional.
Nonetheless, $P^+ p(x_\perp)$ is a state- and frame-independent quantity
that can be cast into units of Pascals,
and accordingly encodes some intrinsic property of the proton,
and may also give some intuitive insight into the rough magnitude of
pressures present inside the proton.
We find with the parameters from Ref.~\cite{Burkert:2021ith}
that $P^+ p(0) = 2.35564\cdot10^{35}$~Pa---the same order of magnitude
suggested by the Breit frame analysis of Ref.~\cite{Burkert:2018bqq}.

\begin{table}[t]
  \caption{
    Root-mean-square radii of the proton on the light front.
    The quoted sources provide a three-dimensional radius that we have
    converted into a 2D light front radius
    using the definitions given in the table.
  }
  \label{tab:radii}
  \setlength{\tabcolsep}{0.3em}
  \renewcommand{\arraystretch}{1.3}
  \begin{tabular}{@{} ccccc @{}} 
    \toprule
    ~ &
    Radius (fm) &
    Uncertainty (fm) &
    Definition &
    Source \\
    \hline
    Pressure & $0.518$ & $0.062_{\mathrm{fit}}+0.126_{\mathrm{sys}}$ &
     Eq.~(\ref{eqn:radius:mechanical}) & This work \\
    Mass     & $0.45$   & $0.02$   & $\sqrt{4A'(0)}$   & Ref.~\cite{Kharzeev:2021qkd} \\
    Axial    & $0.55$   & $0.17$   & $\sqrt{4G_A'(0)}$ & Ref.~\cite{Hill:2017wgb} \\
    Charge   & $0.6266$ & $0.0017$ & $\sqrt{4F_1'(0)}$ & Ref.~\cite{CODATA:2018} \\
    \bottomrule
  \end{tabular}
\end{table}

Using the form in Eq.~(\ref{eqn:Dt}),
there is a simple expression for the mechanical radius, or pressure radius:
\begin{align}
  \langle x_\perp^2(\mathrm{mech}) \rangle
  &=
  \frac{4(\alpha-1)}{\Lambda^2}
  \,.
\end{align}
Our result for the pressure radius is given in Table~\ref{tab:radii},
which also includes several other light front proton radii. 
Note that light front radii differ from the usual
three-dimensional radii defined in the literature,
and are usually just a factor $\sqrt{2/3}$ smaller.
The charge radius, obtained from  the Dirac form factor $F_1$(t),
also differs from the usual Sachs radius
due to relativistic spin effects~\cite{Miller:2007uy}:
\begin{align}
  \langle x_\perp^2({\mathrm{charge}}) \rangle
  =
  \frac{2}{3} r^2_{\mathrm{Sachs}}
  - \frac{\kappa}{M}
  \,,
\end{align}
where $\kappa$ is the proton's anomalous magnetic moment.

Looking at Table~\ref{tab:radii},
the systematic error bars on the pressure radius make any definitive
comparison between it and the other radii difficult.
However, taking the central values seriously
yields  an  apparent ordering of the root-mean-square radii:
\begin{align}
  {x_\perp}({\mathrm{mass}})
  <
  {x_\perp}({\mathrm{pressure}})
  <
  {x_\perp}({\mathrm{axial}})
  <
  {x_\perp}({\mathrm{charge}})
  \,.
\end{align}
Crucially, the apparent spatial extent of the proton differs
depending on how its spatial extent is defined---and the proton
can look bigger or smaller depending on what probe or process is used.
Taking these as strict inequalities cannot be justified with the
uncertainties quoted in Tab.~\ref{tab:radii}.
However, the mass and charge radii are definitively different.
If this ordering roughly holds,
it's worth speculating on what factors might be at play.

To start, the charge radius notoriously obtains a contribution
from the pion cloud~\cite{Thomas:1981vc,Cloet:2012cy,Cloet:2014rja}
that is absent from the axial radius~\cite{Strikman:2009bd},
the latter of which is expected to be smaller for this reason.
By contrast, the pion cloud does carry energy and can reasonably
be expected to exert pressure,
and thus we may expect it to contribute to the mass and pressure radii.

Other factors are likely at play, however.
The ``mass radius'' is actually the radius of the $P^+$
density~\cite{Freese:2019bhb,Freese:2021czn},
and accordingly weighs configurations more strongly when
a single quark carries a large portion of the proton's forward momentum.
These configurations notoriously have small spatial
extent~\cite{Frankfurt:1985cv,Hen:2016kwk},
thus biasing the mass radius towards being small.
The pressure radius may also tend towards being small because
pressure compounds upon itself at greater ``depth,''
i.e., closer to the proton center.
It would be interesting to know with greater certainly
whether the pressure radius really exceeds the mass radius,
and also how it compares to the axial radius.
It would thus be prudent to pursue higher-precision measurements
of DVCS from the proton in order to obtain stronger constraints
on the proton pressure densities and its mechanical radius.

\subsection{Effects of polarization}

It is possible to obtain pressure densities for transversely-polarized
protons within the light front formalism.
Transverse polarization states are given by superpositions of light
front helicity states:
\begin{align}
  | s_T = \mathbf{s}_\perp \rangle
  =
  \frac{
    \left| \lambda = +\frac{1}{2} \right\rangle
    +
    e^{i\phi_s}
    \left| \lambda = -\frac{1}{2} \right\rangle
  }{ \sqrt{2} }
  \,,
\end{align}
and accordingly, expectation values for transverse polarization states
involve helicity-flip matrix elements.

Because of these helicity-flip terms,
proton densities---including the pressure densities---obtain a dependence
on the angle $\phi = \phi_x - \phi_s$
between $\mathbf{x}_\perp$ and $\mathbf{s}_\perp$.
If we use $p(x_\perp)$ to denote the pressure density of a
light front helicity state,
the pressure density for a transversely polarized state is:
\begin{align}
  p_T(x_\perp,\phi)
  &=
  p(x_\perp)
  +
  \frac{\sin\phi}{2M}
  p'(x_\perp)
  \,.
\end{align}
This relation applies to all of the isotropic, radial, and tangential pressures.

\begin{figure}
  \includegraphics[width=\columnwidth]{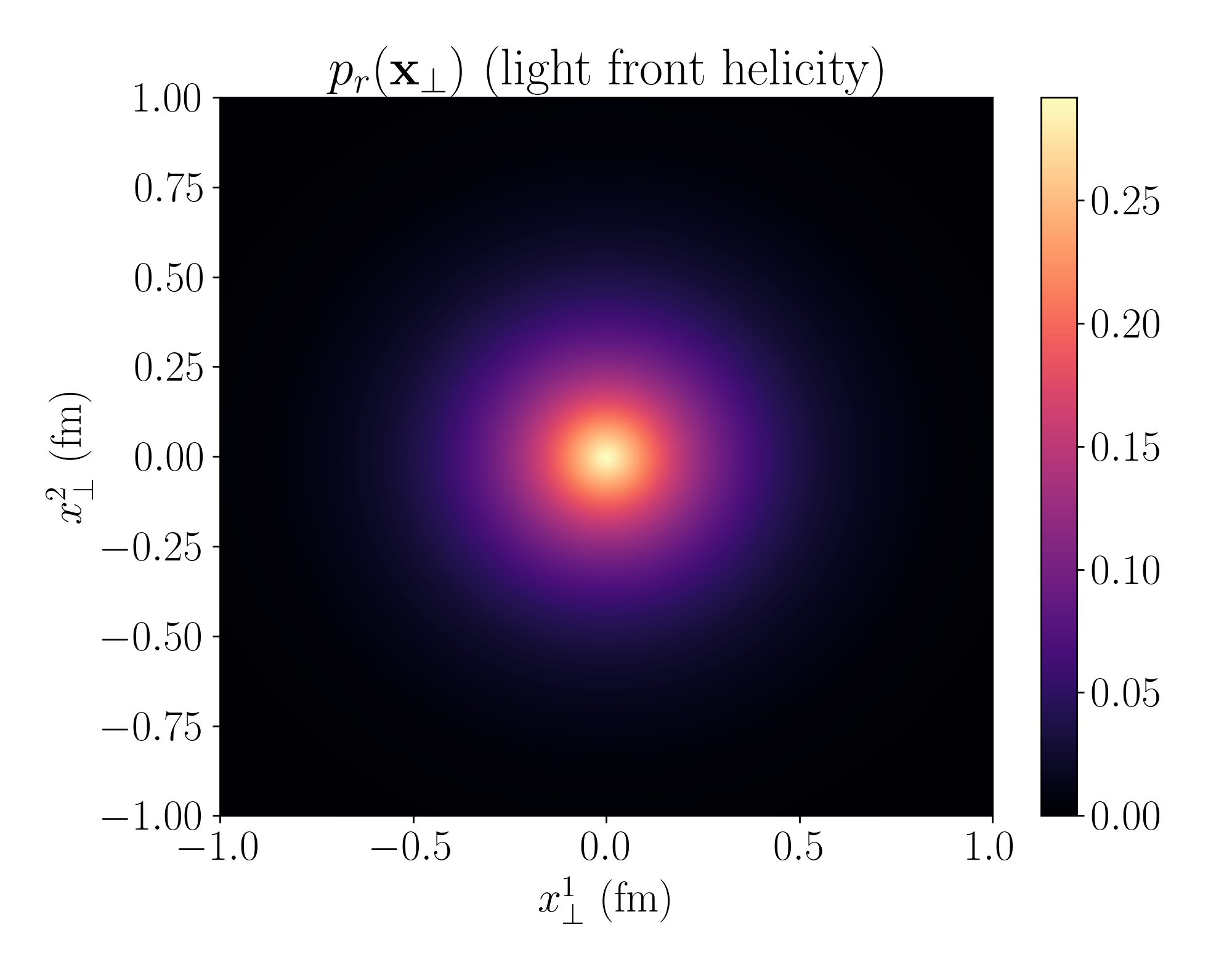}
  \includegraphics[width=\columnwidth]{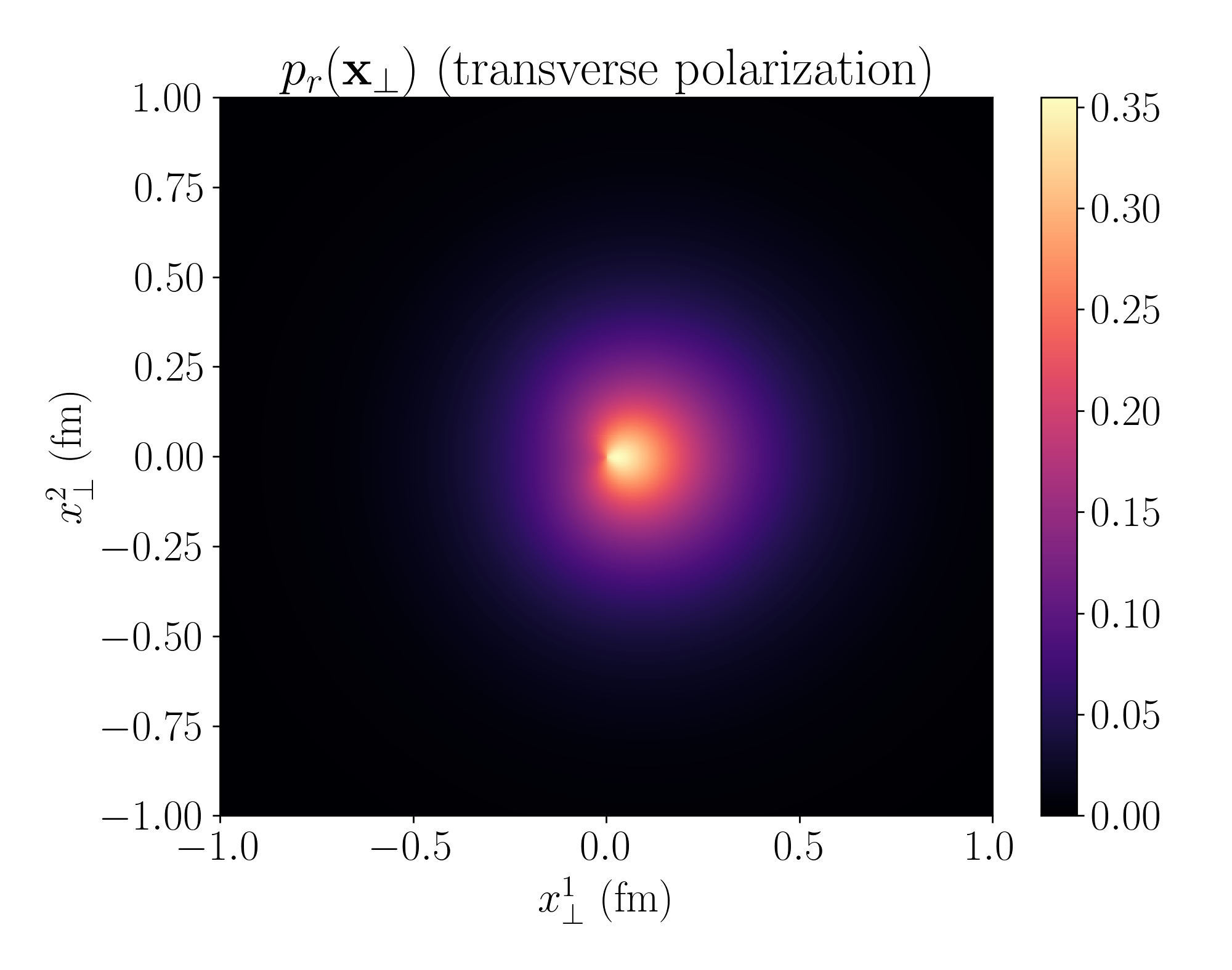}
  \caption{
    2D plots of the radially-directed pressure in the proton.
    A factor $1/P^+$ has been removed from the plotted quantities,
    which are in units units GeV$^2$/fm$^2$.
    The top panel is for longitudinally polarized protons,
    and the bottom panel for transversely-polarized protons
    with the spin in the $+\hat{y}$ direction.
  }
  \label{fig:pressure:2D}
\end{figure}

The 2D radial pressure densities for longitudinally and transversely polarized
protons are plotted in Fig.~\ref{fig:pressure:2D}.
The longitudinally-polarized proton has an azimuthally symmetric pressure.
However, the transversely-polarized proton has a greater concentration of
pressure to the right of ($-90^\circ$ from)
the spin direction.
This finding is reminiscent of a similar finding about electric charge density
in Ref.~\cite{Carlson:2007xd}.

Interestingly, the transverse pressure distribution suggests that---when
analyzed in a light front framework using pressure densities---the
proton is not shaped like a sphere.
This is not too surprising, since the spin axis identifies a particular
direction in space, with respect to which directions such as
right and left can be defined~\cite{Miller:2003sa}.


\section{Discussion and interpretation}

When interpreting the results for the pressures,
it is important to keep in mind their proper physical interpretations.
An especially important fact to bear in mind is that the net force everywhere
in the hadron is identically zero---a statement that the hadron is in
internal equilibrium.

\begin{figure}
  \centering
  \includegraphics[width=0.7\columnwidth]{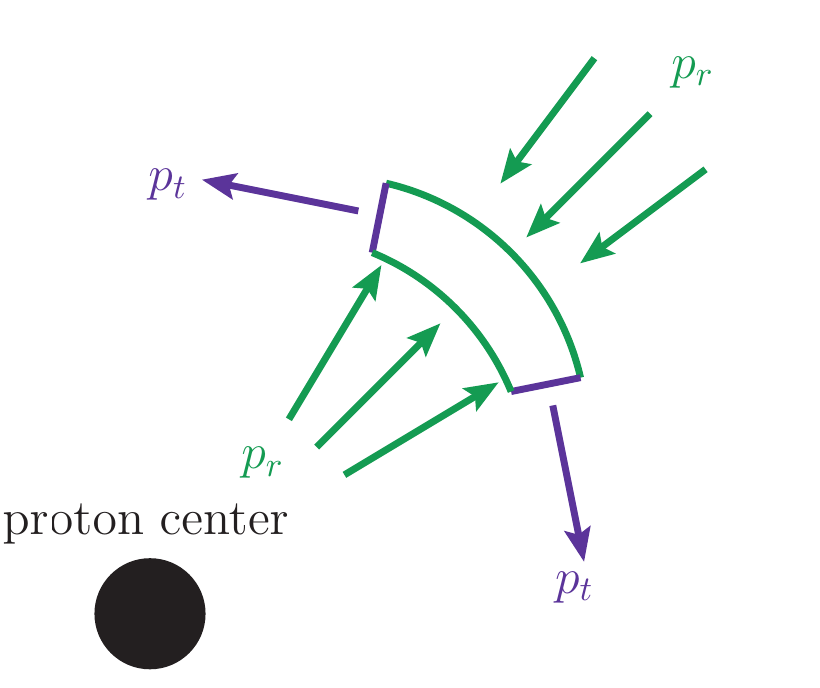}
  \caption{
    A cartoon depicting forces acting on a slab within the proton.
    The net force acting on this slab is zero,
    but forces acting on any side of the slab from the outside
    may be non-zero.
    These non-zero surface forces constitute the pressure
    in the proton.
  }
  \label{fig:slab}
\end{figure}

We clarify the situation further using Fig.~\ref{fig:slab},
which depicts the forces exerted on a small slab within the proton
by the remainder of the proton,
specifically in a case where $p_r > 0$ and $p_t < 0$.
The net force on this piece of the proton is zero,
but there are non-zero forces acting on each side of the slab.
Since $p_r(x_\perp) \geq 0$,
the radially-facing sides of the slab are both pushed on from
the outside.
When $p_t(x_\perp) > 0$,
the tangentially-facing sides are also pushed on,
but when $p_t(x_\perp) < 0$,
these sides are pulled on by the remainder of the proton instead.

As seen in Fig.~\ref{fig:pressure}, the radial pressure in the proton
is strictly positive.
Although attractive and repulsive forces are both present in the photon,
the attractive forces overwhelm the repulsive forces in the radial direction
at all distances.
The balance of forces keeping the proton in equilibrium is thus primarily,
in the radial direction,
between repulsive forces
acting in both the inward and outward radial directions.

On the other hand, at $0.23\pm 0.11~\mathrm{fm} $ from the proton's center,
the tangential pressure changes sign from positive to negative.
This means that at distances less than $0.23$~fm from the proton's center,
the forces in all directions are primarily repulsive,
while at distances greater than $0.23$~fm,
the forces in the $\hat{\phi}$ direction are primarily attractive.
This leads to a scenario where elements of the proton that are far
from its center are being pushed from the radial directions and pulled
around the proton, suggesting a reverse spaghettification.

The isotropic pressure $p(x_\perp)$
averages over the pressures in all directions,
telling us on average whether the majority of forces at a distance
$x_\perp$ from the proton's center are repulsive or attractive.
The pressure crosses zero at ${x_\perp = 0.43\pm0.12~\mathrm{fm} }$,
meaning the forces at shorter distances are primarily repulsive
and forces at longer distances are primarily attractive.
We stress, however, that the forces at these spatial locations are
primarily repulsive or attractive \emph{averaged over directions},
and not towards or away from the proton's center.


\section{Conclusion}

The empirical extraction of $D(t)$ in
Ref.~\cite{Burkert:2021ith}
is used to obtain transverse densities of the isotropic, radial, and tangential pressures
in the proton within the light front formalism.
A physical interpretation of these pressures is provided along with computations of the
empirical mechanical radius associated with them.
Since---in contrast to the picture provided by the Breit frame---transverse
densities on the light front are relativistically correct,
the densities obtained in this work should be interpreted as the genuine
empirical pressure densities of the proton implied by the
findings of Ref.~\cite{Burkert:2021ith}.

\begin{acknowledgments}
  The authors would like to thank Volker Burkert,
  Latifa Elouadrhiri,  F.X.\ Girod
  and M. V. Polyakov for helpful correspondence.
  This work was supported by the U.S.\ Department of Energy
  Office of Science, Office of Nuclear Physics under Award Number
  DE-FG02-97ER-41014.
\end{acknowledgments}


\bibliography{main.bib}

\end{document}